\documentclass{PoS}

\usepackage[cp1250]{inputenc}
\usepackage{latexsym}
\usepackage{subfigure}
\usepackage{multicol}
\usepackage{multirow}
\usepackage{mathrsfs}
\usepackage{color}
\usepackage{verbatim}
\usepackage{amsmath}

\title{Non-perturbative determination of improvement coefficients using coordinate space correlators in $N_f=2+1$ lattice QCD}

\ShortTitle{Non-perturbative determination of improvement coefficients}

\author{\speaker{Piotr Korcyl}\\
        Institut f\"ur Theoretische Physik, Universit\"at Regensburg, D-93040 Regensburg, Germany\\
        E-mail: \email{piotr.korcyl@ur.de}}

\author{Gunnar S. Bali\\
        Institut f\"ur Theoretische Physik, Universit\"at Regensburg, D-93040 Regensburg, Germany\\
        E-mail: \email{gunnar.bali@ur.de}}

\abstract{We determine quark mass dependent order $a$ improvement terms of the form $b_J am$ for non-singlet scalar,
pseudoscalar, vector and axialvector currents, using correlators in coordinate space. We use a set of CLS ensembles
comprising non-perturbatively improved Wilson Fermions and the tree-level Luescher-Weisz gauge action
at $\beta=3.4,3.46,3.55$ and $\beta=3.7$, corresponding to lattice spacings $a$ ranging from $0.05$ fm to $0.09$ fm.
We report the values of the $b_J$ improvement coefficients which are proportional to non-singlet quark
mass combinations and also discuss the possibility of determining the $\bar{b}_J$ coefficients which are proportional
to the trace of the quark mass matrix.}

\FullConference{34th annual International Symposium on Lattice Field Theory\\
		24-30 July 2016\\
		University of Southampton, UK}

\begin{document}

\section{Introduction}

Monte Carlo simulations of Quantum Chromodynamics suffer from discretizations effects. Depending on the chosen discretization of the action
such effects can disappear proportionally to a power of the lattice spacing $a$. Wilson's prescription for
the discrete pure gauge action gives $a^2$ cut-off effects,
whereas the Dirac-Wilson operator introduces corrections linear in the lattice spacing $a$ due to the Wilson term.
One can account for the latter following the Symanzik
improvement programme. In the scaling regime the lattice action can be approximated by its continuum counterpart supplemented by
corrections proportional to positive powers of the lattice spacing,
\begin{equation}
S_{\textrm{QCD}}\big( a(\beta) \big) = S_{\textrm{continuum}} + a S_1 + a^2 S_2 + \dots .
\end{equation}
The so-called $\mathcal{O}(a)$-improvement programme for Wilson fermions consists of adding to $S_{\textrm{QCD}}$
irrelevant operators entering $S_1$
with numerical coefficients chosen in such a way as to remove the entire contribution of that term. Improvement of the action
$S_{\textrm{QCD}}$ requires knowledge of the $c_{\textrm{SW}}$ coefficient and, for non-vanishing quark masses, also of
$b_g$ and $b_m$. Full order $a$ improvement requires not only to improve the action but also operators.
Taking the example of the non-singlet axial current, we usually define
\begin{equation}
%A^{jk, \textrm{I}}_{\mu}(x) = \bar{\psi}_j(x) \gamma_{\mu} \gamma_5 \psi_k(x) + a c_A \partial^{\textrm{sym}}_{\mu} P^{jk}(x)
%\end{equation}
%\textrm{ and }
%\begin{equation}
A^{jk, \textrm{R}}_{\mu}(x) = Z_A \big(1 + a b_A m_{jk}  + a 3 \bar{b}_A \overline{m} \big) \big\{
\bar{\psi}_j(x) \gamma_{\mu} \gamma_5 \psi_k(x) + a c_A \partial^{\textrm{sym}}_{\mu} P^{jk}(x) \big\}
\end{equation}
Hence, for full $\mathcal{O}(a)$-improvement of a non-singlet quark bilinear $J$, using massive Wilson fermions
one needs $c_{\textrm{SW}}$, $b_g$, $b_m$, $c_J$, $b_J$, $\bar{b}_J$. The remaining cut-off effects vanish as $a^2$.

In the present proceedings we summarize results obtained for the
$b_J$ and $\tilde{b}_J$\footnote{
Determining $\bar{b}_J$ requires $b_g$ which is currently unavailable. Instead, we determine $\tilde{b}_J$ which is a
combination of $\bar{b}_J$ and $b_g$
\begin{equation}
\label{eq:redef}
\tilde{b}_J(g^2)=
\bar{b}_J(g^2)+\frac{b_g(g^2)}{N_f}
\left[\frac{\partial\ln Z_J^R(g^2,a\mu)}{\partial g^2}
-\frac{\gamma_J(g^2)}{4\pi\beta(g^2)}\right]g^2\,.
\end{equation}
For practical purposes the knowledge of $\tilde{b}_J$ is sufficient. For a derivation of Eq.~\eqref{eq:redef}
and more details, see Section II of Ref.\ \cite{our}.
} improvement coefficients. For more details we refer the Reader to Ref.~\cite{our}.
We follow the proposal of Ref.~\cite{martinelli}
and use coordinate space correlators to construct observables sensitive to $b_J$ and $\tilde{b}_J$. In particular,
we adapt the original suggestion to the case of the CLS ensembles with $N_f=2+1$ dynamical flavours \cite{cls,rqcd} and implement
several improvements and modifications. We present results for the flavour non-singlet scalar, pseudoscalar, vector
and axialvector currents. The CLS ensembles feature non-perturbatively improved
Wilson fermions and the tree-level L\"uscher-Weisz gauge action.
%The necessary improvement coefficients were estimated: $c_{\textrm{SW}}$ in \cite{csw} and
%$c_A$ in \cite{ca}.
$c_{\textrm{SW}}$ was estimated in \cite{csw}, whereas $c_A$ in \cite{ca}.
We simulate at the bare inverse coupling constant values
$\beta=3.4, 3.46, 3.55$ and $3.7$, which correspond to lattice spacings $a\in[0.05,0.09]\,\textmd{fm}$.
The subset of CLS ensembles used in the present study is presented in table \ref{tab. cls ensembles}.

\begin{table}
\begin{center}
\begin{tabular}{|c||c|c|c|c|c|}
\hline
$\beta$ & name & $\kappa_l$ & $\kappa_s$ & \# conf. & step\\
\hline
3.4 & H101 & 0.136759 & 0.136759 & 100 & 40\\
3.4 & H102 & 0.136865 & 0.136549339 & 100 & 40\\
3.4 & H105 & 0.136970 & 0.13634079 & 103 & 20\\
3.4 & H106 & 0.137016 & 0.136148704 & 57 & 20 \\
3.4 & H107 & 0.136946 & 0.136203165  & 49 & 20\\
3.4 & C101 & 0.137030 & 0.136222041 & 59 & 40\\
3.4 & C102 & 0.137051 & 0.136129063  & 48 & 40\\
3.4 & rqcd17 & 0.1368650 & 0.1368650  & 150 & 40 \\
%3.4 & rqcd21 & P & 32 & 32 & 0.1368130 & 0.1368130 & 21.92774 & 50 & 10 \\
3.4 & rqcd19 & 0.13660  & 0.13660  & 50 & 40 \\
\hline
3.46 & S400 & 0.136984 & 0.136702387  & 83 & 40\\
\hline
3.55 & N203 & 0.137080 & 0.136840284  & 74 & 40\\
\hline
3.7 & J303 & 0.137123 & 0.1367546608  & 38 & 40\\
\hline
\end{tabular}
\caption{List of CLS ensembles used in the study. For details see Ref.~\cite{cls,rqcd}.
The number of independent measurements for each ensemble is given in column 5.
"step" denotes the spacing in MDU between consecutive measurements. Such spacing guarantees that autocorrelations are negligible.
\label{tab. cls ensembles}}
\end{center}
\end{table}

%\subsection{Notation}

For completeness we briefly introduce some notations. We denote quark mass averages as
\begin{equation}
m_{jk} = \frac{1}{2}(m_j + m_k)\,, \qquad m_j = \frac{1}{2a} \left( \frac{1}{\kappa_j} - \frac{1}{\kappa_{\textrm{crit}}} \right), \qquad \overline{m} = \frac{1}{3}\left(m_s + 2 m_{\ell}\right)\,.
\end{equation}
with the subscripts $j,k$ corresponding to different quark content: $j,k= 1,2,3 = \textrm{light}, \textrm{light}, \textrm{strange}$.
The light quark mass $m_{\ell}$ and the average quark mass $\overline{m}$ can be used to parameterize the mass dependence of physical observables.

Our primary observable are the connected Euclidean current-current correlation functions. We denote their continuum,
renormalized in the scheme $R$ at a scale $\mu$, (where, for example, $R=\overline{\mathrm{MS}}$), versions as
\begin{equation}
G_{J^{(jk)}}^R(x,m_{\ell},m_s;\mu)=\left\langle\Omega\left|T\,J^{(jk)}(x)\overline{J}^{(jk)}(0)\right|\Omega\right\rangle^R.
\end{equation}

\section{Method}

\begin{figure}
\begin{center}
\includegraphics[width=0.45\textwidth]{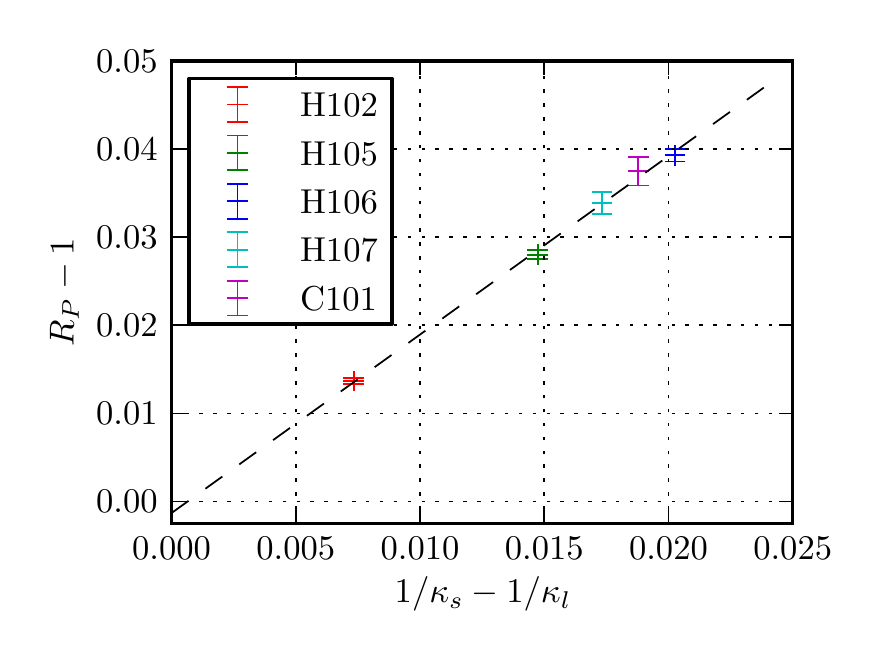}
\includegraphics[width=0.45\textwidth]{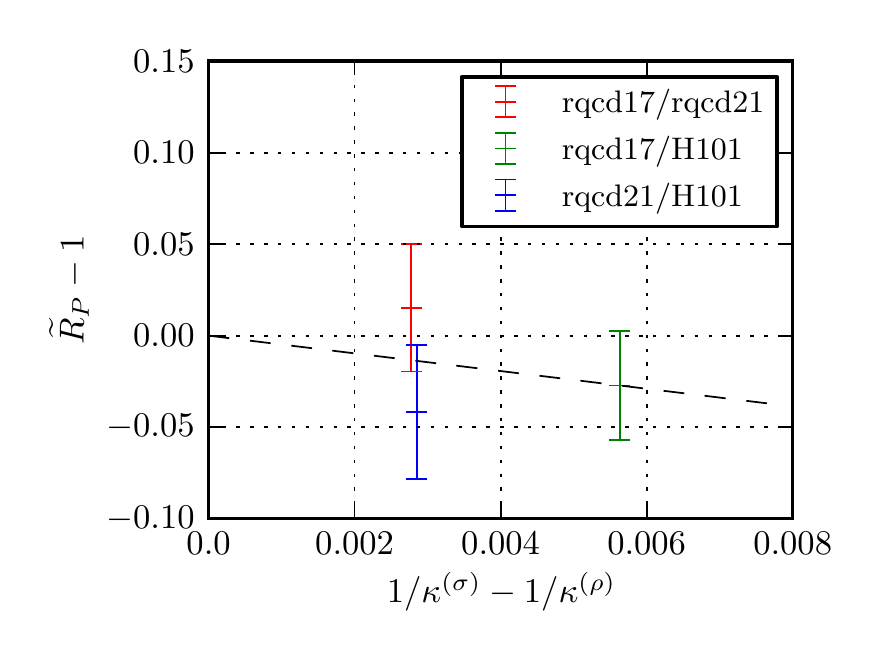}
\caption{Example of data in the pseudoscalar channel for a given vector $n$. The quantity plotted
is $R_P-1$ for $n=(0,1,1,1)$ on the left and $\widetilde{R}_P-1$ for $n=(0,1,2,2)$ on the right
against quark mass difference. A linear dependence is expected, with a slope corresponding
to the improvement coefficient $b_P$ and $\bar{b}_P$. \label{fig. example}
}
\end{center}
\end{figure}

We start with the following two observations valid for correlation functions at distances $x^2 \ll 1/\Lambda^2_{\textrm{QCD}}$
when $m^2 < \Lambda_{\textrm{QCD}}^2$ \cite{martinelli, our}.
\begin{enumerate}
\item The continuum correlation function differs from that of the massless case
by mass dependent terms
\begin{equation}
G_{J^{(jk)}}^R(x,m_{\ell},m_s;\mu)=G_{J^{(jk)}}^R(x,0,0;\mu)
\times
\left[1+\mathcal{O}\left(m^2x^2,m^2\langle FF\rangle x^6,m\langle \overline{\psi}\psi\rangle x^4,m\langle\overline{\psi}\sigma F\psi\rangle x^6\right)\right]\,,
\label{eq.1}
\end{equation}

\item The continuum flavour non-singlet Green function $G^R$ above can be related to
the corresponding Green function $G$ obtained in the lattice scheme
at a lattice spacing $a=a(g^2)$ as follows:
\begin{equation}
G_{J^{(jk)}}^R(x,m_{\ell},m_s;\mu)=\left(Z_J^R\right)^2\!(g^2,a\mu)
\times\left(1+2b_Jam_{jk}+6\tilde{b}_Ja\overline{m}\right)
G_{J^{(jk),I}}(n,am_{jk},a\overline{m};g^2)\,,
\label{eq.2}
\end{equation}
\end{enumerate}
Note that we replaced $Z^R(\tilde{g}^2)$ by $Z^R(g^2)$ and absorbed the difference, replacing $\bar{b}_J$ by $\tilde{b}_J$,
see Eq.~\eqref{eq:redef}.
Hence, by constructing a ratio of two correlation functions with identical Dirac structure we can eliminate the massless
correlation function on the right-hand side of Eq.~\eqref{eq.1} and the mass independent renormalization constant in
Eq.~\eqref{eq.2}. We arrive at the following expression
\begin{multline}
\frac{G_{J^{(jk)}}\left(n,am^{(\rho)}_{jk},a\overline{m}^{(\rho)};g^2\right)}
{G_{J^{(rs)}}\left(n,am^{(\sigma)}_{rs},a\overline{m}^{(\sigma)};g^2\right)}
=
1+2b_Ja\left(m^{(\sigma)}_{rs}-m^{(\rho)}_{jk}\right) +6\tilde{b}_Ja\left(\overline{m}^{(\sigma)}-\overline{m}^{(\rho)}\right)
+\mathcal{O}\left(a^2,x^2\right),
\end{multline}
where the $\rho$ and $\sigma$ indices distinguish different points in the $\kappa_{\ell} - \kappa_s$ parameter plane.
We now define two observables which are directly proportional to the improvement coefficients we are after. In the
first case we choose one ensemble away from the symmetric line, i.e. $m_{\ell} \ne m_s$, and measure two correlation functions: one with two light quarks and the second
with a light and a strange quark. Their ratio is approximated by
\begin{equation}
R_J(x, m_{12}^{(\rho)}, m_{13}^{(\rho)}) \equiv \frac{G_{J^{(12)}}\left(n,am^{(\rho)}_{12},a\overline{m}^{(\rho)};g^2\right)}
{G_{J^{(13)}}\left(n,am^{(\rho)}_{13},a\overline{m}^{(\rho)};g^2\right)} =
1 + 2 b_J a (m_{13}^{(\rho)} - m_{12}^{(\rho)}) %\left(m_{13}^{(\rho)} - m_{12}^{(\rho)}\right)
= 1 + 2 b_J a \delta m
\label{eq. r}
%&=1+b_J\left(\frac{1}{\kappa_s}-\frac{1}{\kappa_{\ell}}\right)\,.
\end{equation}
Note that $\kappa_{\textrm{crit}}$ as well as $\tilde{b}_J$ cancel. For the second observable we choose
two ensembles on the symmetric line and we define $\widetilde{R}_J(x, \delta \overline{m}=\overline{m}^{(\sigma)} - \overline{m}^{(\rho)})$
\begin{equation}
\widetilde{R}_J(x, \overline{m}^{(\sigma)}, \overline{m}^{(\rho)}) \equiv \frac{G_{J^{(12)}}
\left(n,a\overline{m}^{(\rho)},a\overline{m}^{(\rho)};g^2\right)}
{G_{J^{(12)}}\left(n,a\overline{m}^{(\sigma)},a\overline{m}^{(\sigma)};g^2\right)}
=1+(2b_J+6\tilde{b}_J)a (\overline{m}^{(\sigma)} - \overline{m}^{(\rho)}) = \widetilde{R}_J(x, \delta \overline{m}),
%=1+(2b_J+6\tilde{b}_J)a \delta \overline{m},
\label{eq. rtilde}
%\left(\overline{m}^{(\sigma)}-\overline{m}^{(\rho)}\right)\nonumber
%&=
%1+(b_J+3\tilde{b}_J)\left(\frac{1}{\kappa^{(\sigma)}}-\frac{1}{\kappa^{(\rho)}}\right).
\end{equation}
which gives us sensitivity to $\tilde{b}_J$ once $b_J$ is known. Again, no knowledge of $\kappa_{\textrm{crit}}$ is needed.
Figure \ref{fig. example} shows $R_J(x, \delta m)$ and $\widetilde{R}_J(x, \delta \overline{m})$ as functions
of $\delta m$ and $\delta \overline{m}$ for a given vector $n$. A linear behaviour of the data confirms the validity of expressions
Eq.~\eqref{eq. r} and Eq.~\eqref{eq. rtilde}.

\section{Short and medium distance corrections}

The observables $R_J(x, \delta m)$ and $\widetilde{R}_J(x, \delta \overline{m})$ %defined in Eq.\ref{eq. r} and Eq.\ref{eq. rtilde}
are affected by undesired cut-off effects which are mostly visible at short distances ($< 0.2$ fm) as well as mass dependent non-perturbative corrections relevant
at medium distances ($\approx 0.3-0.4$ fm). We compensate
for both of them at tree-level perturbation theory and propose improved observables
$B_J(x, \delta m)$ and $\widetilde{B}_J(x, \delta \overline{m})$ with these effects subtracted.

\begin{figure}
\begin{center}
\includegraphics[width=0.45\textwidth]{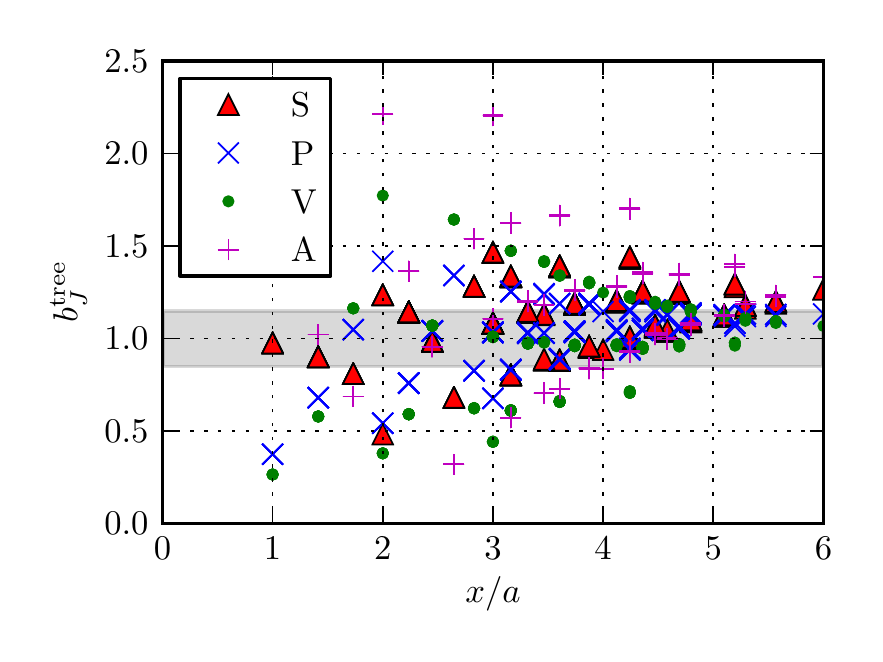}
\includegraphics[width=0.45\textwidth]{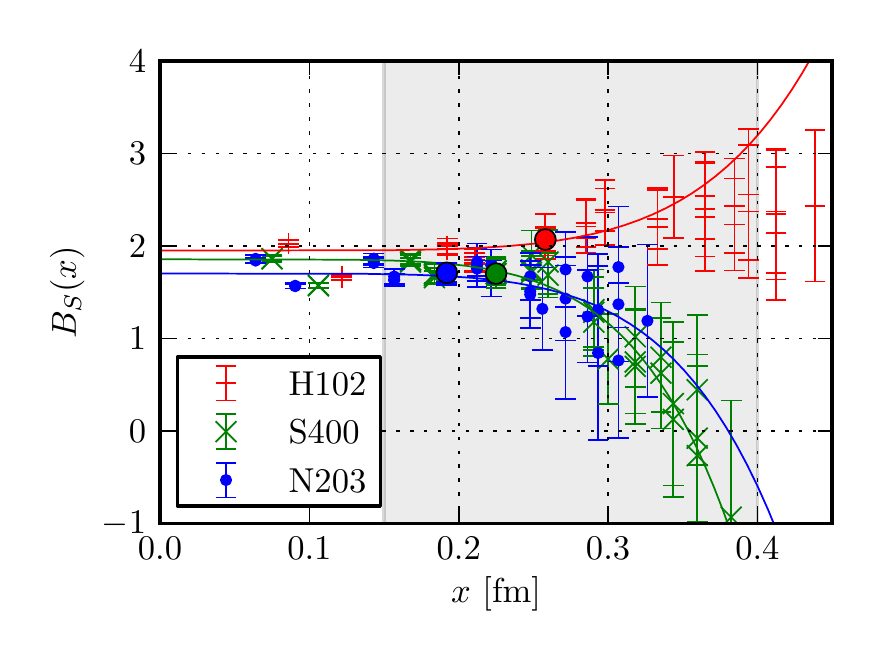}
\caption{Left panel: Observables $R_J$ estimated in tree-level lattice perturbation theory plotted
against the expected continuum prediction $b^{\textrm{tree}}_J=1$. In the analysis we use only vectors for which
the tree-level cut-off artifacts are smaller than $15\%$, i.e. which lie within the shaded band.
Right panel: Data for $b_S$ for all investigated vectors $n$ for three ensembles: H102, S400 and N203. The shaded area corresponds to the
interval of $x$ used in the fit which estimates the size of $x^6$ corrections. \label{fig. data}}
\end{center}
\end{figure}

Figure \ref{fig. data} shows the observable $R_J(x, \delta m)$ computed in
tree-level lattice perturbation theory. The continuum tree-level prediction is $b^{\textrm{tree}}_J=1$, hence all deviations
from that value on figure \ref{fig. data} are due to non-leading lattice artifacts. Using these data we can construct a set of distances $x$ for which cut-off
effects are minimal. Subsequently, we only consider vectors
for which the tree-level cut-off effects are smaller than $15\%$. The improved observables $B_J$ and $\widetilde{B}_J$ have these
corrections subtracted.

At medium distances we can predict the behaviour of our data as a function of $x$ using the Operator Product Expansion.
For the ratio of correlation functions one gets
\begin{equation}
\frac{G_{J^{(12)}}(x)}
{G_{J^{(34)}}(x)}=1+(A^J_{12}-A^J_{34})x^2 +\left[\left(A^J_{34}\right)^2-A^J_{12}A^J_{34} +B^J_{12}-B^J_{34}\right]x^4+\cdots,
\end{equation}
with the mass dependent coefficients
\begin{equation}
A^J_{jk}=-\frac14\left(m_j^2+m_k^2+\frac{m_jm_k}{s_J}\right)\,, \qquad
B^J_{jk}=\frac{\pi^2}{32N}\langle FF\rangle+\frac{m_j^2m_k^2}{16}+\frac{\pi^2}{8N}\frac{2+s_J}{s_J}(m_j+m_k)\langle\overline{\psi}\psi\rangle\,.
\end{equation}
where $s_J$ are constants and depend on the Dirac structure, $s_S = 1 = - s_P$ and $s_A = \frac{1}{2} = - s_V$.
Knowing the
explicit form of these corrections and using the Gell-Mann-Oakes-Renner relation
%\begin{equation}
$(m_j + m_k)\langle \bar{\psi} \psi \rangle = - F_0^2 M^2_{jk},$
%\end{equation}
where
$M_{jk}$ denotes the mass of a pseudoscalar meson composed of
(anti)quarks of masses $m_j$ and $m_k$ and the pion decay
constant in the $N_f = 3$ chiral limit reads $F_0 = 86.5(1.2)$ MeV \cite{flag, pdg}, we
correct our observables $R_{J}$ and $\widetilde{R}_J$ by subtracting the leading
continuum corrections.

Hence, we arrive at the improved observable
\begin{align}
B_J(x, \delta m )&\equiv\biggl[R_J(x, \delta m )-R^{\mathrm{tree}}_J(x, \delta m )
\left.+\frac{\pi^2}{8N}\frac{2+s_J}{s_J}\left(M_{\pi}^2-M_K^2\right)F_0^2x^4\right]\times\left(\frac{1}{\kappa_s}-\frac{1}{\kappa_{\ell}}\right)^{-1}  \nonumber \\
&=b_J
+\mathcal{O}(x^6)+\mathcal{O}(g^2 a^2)+\cdots\,,
\end{align}
and a similar expression for $\widetilde{B}_J(x, \delta \overline{m} )$.
%\begin{align}
%\widetilde{B}_J(x, \delta \overline{m} )&\equiv\biggl[\widetilde{R}_J(x, \delta \overline{m} )
%-\widetilde{R}^{\mathrm{tree}}_J(x,
%\delta \overline{m} )
%\left.+\frac{\pi^2}{8N}\frac{2+s_J}{s_J}\left( \delta {M_{\pi}}^2 \right)F_0^2x^4\right]\times\left(\frac{1}{\kappa^{(\sigma)}}-\frac{1}{\kappa^{(\rho)}}\right)^{-1} \nonumber \\
%&=b_J+3\tilde{b}_J
%+\mathcal{O}(x^6)+\mathcal{O}(g^2 a^2)+\cdots\,.
%\end{align}

Figure \ref{fig. data} shows the numerical data for the improved observable in the scalar channel $B_S$. We see that the data
are rather flat up until $\approx 0.3$ fm. The remaining curvature is due to higher order corrections. We take them into account by
attributing a systematic uncertainty to our results which we define as the size of the contribution proportional to $x^6$.
We estimate the latter by fitting to a function containing a term $x^6$ within an interval of $0.15 - 0.4$ fm.

%\begin{figure}
%\begin{center}
%\includegraphics[width=0.6\textwidth]{../ope_test_new_SS_three_phys.pdf}
%\caption{Data for $b_S$ for all investigated vectors $n$ for three ensembles: H102, S400 and N203. The shaded area corresponds to the
%interval of $x$ used in the fit which estimates the size of $x^6$ corrections. \label{fig. data}}
%\end{center}
%\end{figure}

\begin{figure}
\begin{center}
\includegraphics[width=0.45\textwidth]{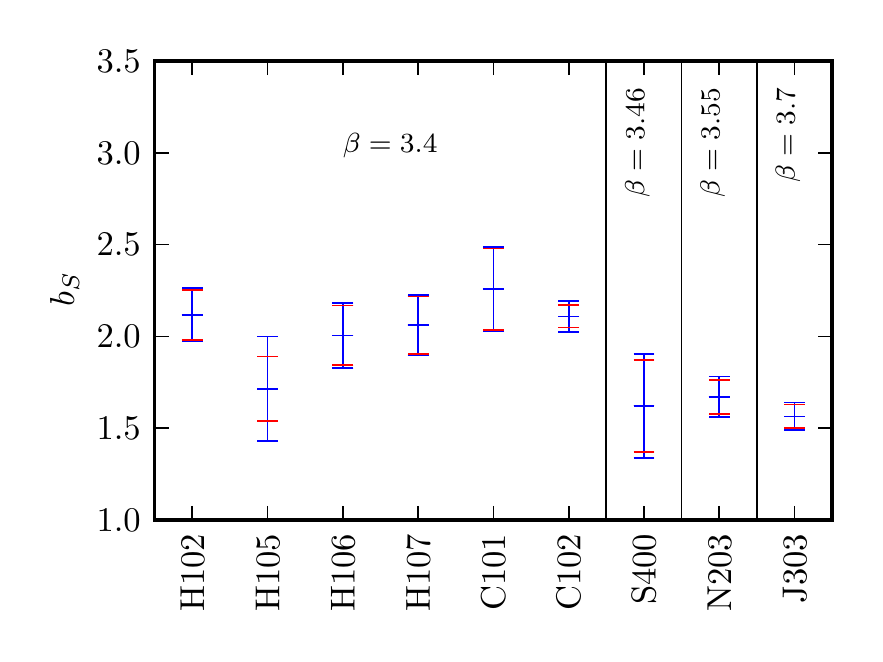}
\includegraphics[width=0.45\textwidth]{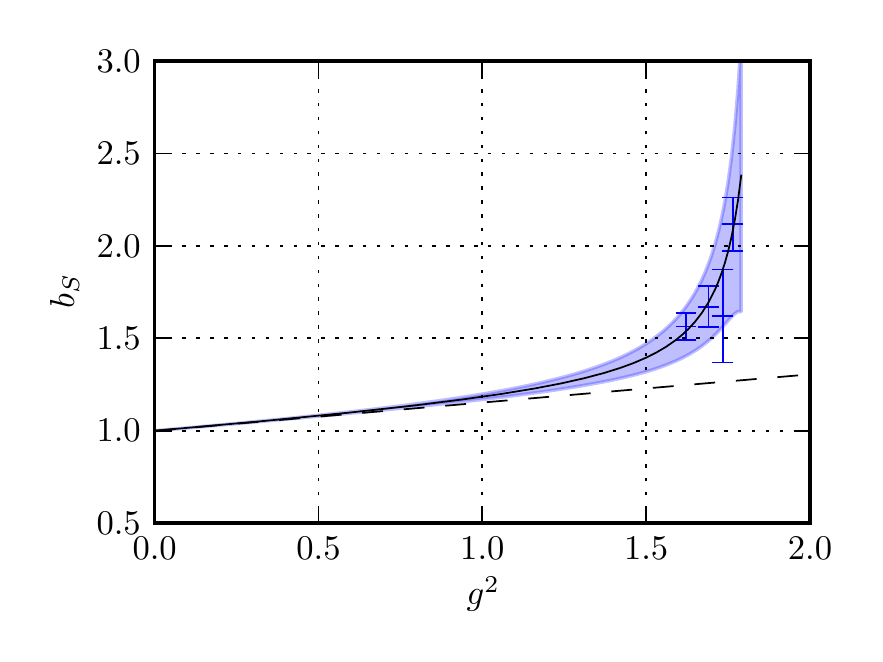}
\caption{Left panel: Results for $b_S$ for all ensembles. Blue error bars correspond to the total uncertainty with the combined
systematic uncertainties of a 20\% uncertainty of the Wilson coefficient of the non-perturbative $x^4$ correction and
an estimate of the size of order $x^6$ corrections. Red error bars correspond to the statistical uncertainty only.
Final results are computed using the vector $n=(0,1,2,2)$ and all its symmetric equivalents.
Right panel: Rational parametrization of $b_S$ as a function of the coupling constant $g^2$. The dashed line is the
one-loop perturbative expectation. \label{fig. results}}
\end{center}
\end{figure}

Results in the scalar channel from all investigated ensembles are summarized in figure \ref{fig. results}. We note that within
one value of the coupling constant the results are compatible within their errors and that the value of the $b_S$ improvement
coefficient systematically decreases with a decreasing coupling constant, i.e. increasing $\beta$.

\vspace{-1.0cm}

\section{Rational parameterizations}

We present our final results for $b_J$ in the form of rational parameterizations. We choose four ensembles: H102, S400, N203 and J303 at four values
of the coupling constant and perform a fit with the following ansatz
%\begin{equation}
$b_J(g^2) = 1 + b^{\textrm{one-loop}}_J g^2 \big(1 + \gamma_J g^2 \big) \big(1 + \delta_J g^2 \big)^{-1},$
%b_J(g^2) &= 1 + b^{\textrm{one-loop}}_J g^2 \big(1 + \gamma_J g^2 \big) \textrm{ for } J = V
%\end{equation}
with $\delta_V = 0$. The final expressions read
\begin{equation}
b_S(g^2) = 1 + 0.11444(1)C_F g^2 \big(1 - 0.439(50) g^2 \big) \big(1 - 0.535(14) g^2 \big)^{-1}, \\
\end{equation}
\begin{equation}
b_P(g^2) = 1 + 0.0890(1)C_F g^2 \big(1 - 0.354(54) g^2 \big) \big(1 - 0.540(11) g^2 \big)^{-1}, \\
\end{equation}
\begin{equation}
b_V(g^2) = 1 + 0.0886(1)C_F g^2 \big(1 + 0.596(111) g^2 \big), \\
\end{equation}
\begin{equation}
b_A(g^2) = 1 + 0.0881(1)C_F g^2 \big(1 - 0.523(33) g^2 \big) \big(1 - 0.554(10) g^2 \big)^{-1},
\end{equation}
and the fit for the case of the scalar channel is shown in figure \ref{fig. results}. The one-loop values of the
improvement coefficients for the discussed lattice action can be obtained from the results of Ref.\cite{oneloop}.

%\begin{figure}
%\begin{center}
%\includegraphics[width=0.6\textwidth]{../test_SS_g0c.pdf}
%\caption{Rational parametrization of $b_S$ as a function of the coupling constant $g^2$. \label{fig. pade}}
%\end{center}
%\end{figure}

\section{$\tilde{b}_J$ improvement coefficients}

At $\beta=3.4$ several ensembles from the symmetric line are available. We used them to test our method to extract
the $\tilde{b}_J$ improvement coefficients. Since the ratio is composed of correlation functions measured on two different
ensembles the statistical noise does not cancel and the statistics used so far yield very imprecise results:
\begin{equation}
\tilde{b}_S =  2.0  (1.3) (0.3), \
\tilde{b}_P =  -3.4  (1.3) (0.6), \
\tilde{b}_V =  -0.1  (0.4) (0.1), \
\tilde{b}_A =   1.4  (0.4) (0.9),
\end{equation}
the first uncertainty being systematic and the second statistical. Several ways are being tested in order to improve the precision, including stochastic multiple point sources similar to the ones
used in Ref.~\cite{hqet} or the Truncated Solver Method of Ref.~\cite{truncated}.

\section{Conclusions}

We implemented and tested a coordinate space method to determine improvement coefficients accompanying quark mass dependent terms.
The method is general
and allows for the determination of improvement coefficients for all quark bilinears. With a negligible numerical effort we
can achieve a 5\% -- 10\% precision on $b_J$. Improvement coefficients proportional to the trace of the mass matrix are accessible,
but need a better statistical precision. These promising results encourage further steps such as the determination of the improvement
coefficients $c_J$ or improvement coefficients for singlet fermion bilinear operators,
which are of relevance in studies of the structure of nucleons, or of
more complicated currents with derivatives.


\begin{thebibliography}{99}
\bibitem{our} P. Korcyl, G. Bali, (2016), arXiv:1607.07090 [hep-lat].
\bibitem{martinelli} G. Martinelli \emph{et al.}, Phys. Lett. B 411, 141 (1997).
\bibitem{cls} M. Bruno \emph{et al.} (CLS),
%\emph{Simulation of QCD with
%𝑁𝑓 = 2 + 1 flavors of non-perturbatively improved Wilson
%Fermions,}
J. High Energy Phys. 02, 043 (2015), arXiv:1411.3982 [hep-lat].
\bibitem{rqcd} G. Bali, E. E. Scholz, J. Simeth, W. S\"oldner, (2016), arXiv:1606.09039 [hep-lat].
\bibitem{csw} J. Bulava, S. Schaefer, %“Improvement of 𝑁𝑓 =
%3 Lattice QCD with Wilson Fermions and tree-level improved
%gauge action,”
Nucl. Phys. B 874, 188 (2013), arXiv:1304.7093 [hep-lat].
\bibitem{ca} J. Bulava, M. Della Morte, J. Heitger, C. Wittemeier (ALPHA Collaboration),
%“Nonperturbative
%improvement of the axial current in 𝑁𝑓 = 3
%Lattice QCD with Wilson Fermions and tree-level improved
%gauge action,”
Nucl. Phys. B 896, 555 (2015), arXiv:1502.04999 [hep-lat].
\bibitem{flag} Sinya Aoki \emph{et al.} (FLAG),
%“Review of lattice results
%concerning low-energy particle physics,”
(2016), arXiv:1607.00299 [hep-lat].
\bibitem{pdg} Keith A. Olive \emph{et al.} (Particle Data Group),
%“Review of
%Particle Physics,”
Chin. Phys. C 38, 090001 (2014).
\bibitem{oneloop}
Y. Taniguchi, A. Ukawa, %gPerturbative calculation
%of improvement coefficients to ..(..2..) for bilinear
%quark operators in Lattice QCD,h
Phys. Rev. D 58,
114503 (1998), arXiv:hep-lat/9806015 [hep-lat].
\bibitem{hqet} P. Korcyl, C. Lehner, T. Ishikawa,
%Non-perturbative renormalization of the static quark theory in a large volume
PoS LATTICE2015 (2016) 254, arXiv:1512.00069 [hep-lat].
\bibitem{truncated} G. Bali, S. Collins, A. Schaefer,
Comput.Phys.Commun.181:1570-1583 (2010), arXiv:0910.3970 [hep-lat]

\end{thebibliography}
\end{document}